\documentclass[reprint, amsmath, amssymb, aps, floatfix]{revtex4-2}

\usepackage{graphicx}
\usepackage[colorlinks=true, urlcolor=blue, citecolor=blue, linkcolor=blue]{hyperref}
\usepackage{cleveref}
\usepackage{bm}
\usepackage{newtxtext, newtxmath}

\newcommand{\rmd}{\mathrm{d}}
\newcommand{\cl}{\mathrm{cl}}
\newcommand{\q}{\mathrm{q}}
\newcommand{\abs}[1]{{\left| {#1} \right|}}
\newcommand{\moment}{\widetilde{\mathcal{M}}}
\newcommand{\avg}[1]{{\left\langle {#1} \right\rangle}}
\newcommand{\mf}{\hat{\mu}}
\newcommand{\micron}{\mu\mathrm{m}}
\renewcommand{\vec}{\bm}
\newcommand{\edit}[1]{{#1}}

\begin{document}

\title{Analytical solutions for quantum radiation reaction in high-intensity lasers}

\author{T. G. Blackburn}
\email{tom.blackburn@physics.gu.se}
\affiliation{Department of Physics, University of Gothenburg, SE-41296 Gothenburg, Sweden}


\begin{abstract}
While the Landau-Lifshitz equation, which describes classical radiation reaction, can be solved exactly and analytically for a charged particle accelerated by a plane electromagnetic wave, no such solutions are available for quantum radiation reaction \edit{(the recoil arising from the successive, incoherent emission of hard photons)}.
Yet upcoming experiments with ultrarelativistic electron beams and high-intensity lasers will explore the regime where both radiation-reaction and quantum effects are important.
Here we present analytical solutions for the mean and variance of the energy distribution of an electron beam that collides with a pulsed plane electromagnetic wave, which are obtained by means of a perturbative expansion in the quantum parameter $\chi_0$.
These solutions capture both the quantum reduction in the radiated power and stochastic broadening, and are shown to be accurate across the range of experimentally relevant collision parameters, i.e. GeV-class electron beams and laser amplitudes $a_0 \lesssim 200$.
\end{abstract}

\maketitle

\section{Introduction}

The electromagnetic fields produced by focused high-power lasers are so strong that the dynamics of relativistic particles enters the regime of strong-field QED~\cite{dipiazza.rmp.2012,gonoskov.rmp.2022,fedotov.pr.2023}.
One process that has attracted much interest is quantum radiation reaction, i.e. the accumulated recoil from the emission of individual high-energy photons~\cite{dipiazza.prl.2010}, which can be as significant to the particle and plasma dynamics as the acceleration induced by the background electromagnetic field~\cite{gonoskov.rmp.2022}.
Experiments with high-intensity lasers have already shown evidence of radiation reaction~\cite{cole.prx.2018,poder.prx.2018} and investigation of strong-field QED effects, including quantum radiation reaction, is a key part of the science case for upcoming and planned laser facilities~\cite{weber.mre.2017,gales.rpp.2018,maksimchuk.2019,turner.epjd.2022,ji.2022,mp3.report}.

Quantum radiation reaction has many possible experimental signatures, including stochastic broadening~\cite{neitz.prl.2013}, straggling~\cite{blackburn.prl.2014}, quenching~\cite{harvey.prl.2017}, and increased angular divergence~\cite{green.prl.2014,vranic.njp.2016}, all of which arise because photon emission is inherently probabilistic.
These works rely largely on the results of numerical simulations, as the theory for quantum radiation reaction is not generally amenable to analytical solution.
By contrast, the Landau-Lifshitz equation~\cite{landau.lifshitz}, which describes classical radiation reaction, can be solved exactly for a general plane-wave background~\cite{dipiazza.lmp.2008} (see also Refs.~\citenum{heintzmann.zpa.1972,hadad.prd.2010,harvey.prd.2011}).
It would be helpful for guidance of future experiments to have analytical solutions that apply in the quantum regime.
In this work we consider the radiation reaction of an ultrarelativistic electron beam in an intense laser background, but note that similar phenomena can be explored with aligned crystals~\cite{wistisen.ncomms.2018,nielsen.prd.2020,dipiazza.plb.2017,khokonov.plb.2019}.

Consider a beam of ultrarelativistic electrons, which has a distribution of Lorentz factors $\gamma$, $\frac{\rmd N_e}{\rmd \gamma}$, characterized by a mean $\mu = \avg{\gamma}$, variance $\sigma^2 = \avg{(\gamma - \mu)^2}$, and other higher order moments including $\varsigma^3 = \avg{(\gamma - \mu)^3}$ and $\kappa^4 = \avg{(\gamma - \mu)^4}$, which are related to the skewness and kurtosis respectively.
This beam collides with a intense laser pulse, which is modelled as a plane electromagnetic wave with angular frequency $\omega_0$ and normalized amplitude $a_0$, such that the electric field as a function of phase $\phi$ is $\vec{E}(\phi) = m \omega_0 a_0 \vec{f}(\phi) / e$.
Here $e$ and $m$ are the elementary charge and electron mass, respectively, and we work in natural units where $\hbar = c  = 1$.
As the electron beam propagates through the laser pulse, it emits radiation and decelerates.

\edit{If the electrons are ultrarelativistic, radiation emission and reaction may be treated within the semiclassical framework proposed by \citet{baier.jetp.1968}.}
Provided that $\gamma \gg a_0$, and $a_0$ is large enough that the locally constant field approximation holds~\cite{ritus.jslr.1985,dinu.prl.2016,dipiazza.pra.2018,blackburn.pop.2018}, the evolution of the mean and variance of the energy distribution is given by~\citep{ridgers.jpp.2017,niel.pre.2018}
    \begin{equation}
    \frac{\rmd \mu}{\rmd \phi} =
        -\frac{2 R_c}{3 \mu_0} \abs{f(\phi)}^2
        \avg{ \gamma^2 g(\chi) },
    \label{eq:MuEvolution}
    \end{equation}
and
    \begin{multline}
    \frac{\rmd \sigma^2}{\rmd \phi} =
        -\frac{4 R_c}{3 \mu_0} \abs{f(\phi)}^2
        \avg{ (\gamma - \mu) \gamma^2 g(\chi) }
    \\
        + \frac{55 R_c \chi_0}{24 \sqrt{3} \mu_0^2} \abs{f(\phi)}^3
        \avg{ \gamma^4 g_2(\chi) },
    \label{eq:SigmaEvolution}
    \end{multline}
where $R_c = \alpha a_0 \chi_0$ is the classical radiation reaction parameter, $\chi_0 = 2 a_0 \mu_0 \omega_0 / m$ is the quantum parameter, and $\mu_0$ is the initial value of the mean.
The unsubscripted $\chi$ appearing in these equations is the instantaneous value of the quantum parameter, $\chi = 2 a_0 \gamma \omega_0 \abs{f(\phi)} / m$, which depends on the instantaneous $\gamma$ and field amplitude.
The two functions $g(\chi)$ and $g_2(\chi)$ describe the role of quantum corrections to radiation reaction and are discussed in \cref{sec:Moments}.

The purpose of this work is to find analytical predictions of the mean and variance in the regime where quantum effects are important, but not dominant.
Equivalent results for the classical regime $\chi_0 = 0$ have been obtained by \citet{neitz.prl.2013} and \citet{vranic.prl.2014,vranic.njp.2016}.
This analysis is extended to the quantum regime and to the whole hierarchy of moments by \citet{niel.pre.2018}.
The dynamics of the energy distribution itself, rather than its moments, under quantum radiation reaction is treated analytically in \citet{bulanov.2023}.
Furthermore, the evolution of the mean and variance in a constant field has been obtained by \citet{torgrimsson.prl.2021,torgrimsson.2023}, using a resummation approach.
The strategy here is to solve \cref{eq:MuEvolution,eq:SigmaEvolution} perturbatively in the small parameter $\chi_0$.
Additionally, to break the infinite hierarchy that arises because the evolution of a given moment depends
on higher-order moments, we make the approximation that that successive moments are smaller than each other, i.e. $\mu \gg \sigma \gg \kappa$.
We begin by discussing the functions $g(\chi)$ and $g_2(\chi)$, then present analytical solutions for the mean and variance of the distribution.

\section{Quantum corrections}
\label{sec:Moments}

    \begin{table}
    \begin{ruledtabular}
    \begin{tabular}{llll}
    $n$	& $\mathcal{M}_\cl(n,\chi)$ 						& $\moment(n,\chi)$ at $\chi \ll 1$									& $\moment(n,\chi)$ at $\chi \gg 1$ \\ \hline
    0		& $\frac{5}{2\sqrt{3}} \alpha m \chi$ 						& $1 - \frac{8}{5\sqrt{3}}\chi + \frac{7}{2}\chi^2$			& $\frac{28\Gamma(2/3)}{3^{5/6} 15} \chi^{-1/3}$ \\
    1		& $\frac{2}{3} \alpha m \chi^3$	& $1 - \frac{55\sqrt{3}}{16}\chi + 48\chi^2$						& $\frac{128\pi}{3^{5/6} 243 \Gamma(7/3)} \chi^{-4/3}$ \\
    2		& $\frac{55}{24\sqrt{3}} \alpha m \chi^5$ 			& $1 - \frac{448\sqrt{3}}{55}\chi + \frac{777}{4}\chi^2$	& $\frac{236\Gamma(5/3)}{3^{5/6} 495} \chi^{-7/3}$
    \end{tabular}
    \end{ruledtabular}
    \caption{Moments of the classical and quantum photon emission rates.}
    \label{tbl:Results}
    \end{table}

Quantum effects are manifest in the two functions $g(\chi)$ and $g_2(\chi)$, which relate moments of the quantum and classical synchrotron emissivities.
In particular, $g(\chi)$ represents the reduction in the radiation power caused by quantum corrections to the synchrotron spectrum.

We will define the $n$th moment of the radiation spectrum to be
	\begin{equation}
	\mathcal{M} (n, \chi) = \int \! (\chi s) ^{n} \frac{\rmd W_\gamma}{\rmd s} \,\rmd s,
	\end{equation}
where $\frac{\rmd W_\gamma}{\rmd s}$ is the photon emission rate per unit proper time, per unit photon normalized energy $s = \omega' / (\gamma m)$, as calculated in the locally constant field approximation~\cite{erber.rmp.1966,ritus.jslr.1985}:
    \begin{multline}
    \frac{\rmd W_\gamma}{\rmd s} =
        \frac{\alpha m}{\sqrt{3} \pi} \left[
        \left( 1 - s + \frac{1}{1-s} \right) K_{2/3}(\xi)
    \right. \\ \left.
        - \int_\xi^\infty \! K_{1/3}(t) \, \rmd t
        \right],
    \end{multline}
where $\xi = 2 s / [3 \chi (1 - s)]$ and $K_n$ is a modified Bessel function of the second kind.
The classical emission rate is obtained by replacing $1-s \to 1$ wherever it appears.
The zeroth moment is the total emission rate $\mathcal{M}(0, \chi) = W_\gamma$.
The normalized $n$th moment is
	\begin{equation}
	\moment (n, \chi) = \frac{\mathcal{M}_\q (n, \chi)}{\mathcal{M}_\cl (n, \chi)}.
	\end{equation}
The subscripts denote whether the quantum or classical emission rates are to be used when evaluating the integrals.
For example, the quantum correction to the radiated power~\cite{erber.rmp.1966}, sometimes called the \emph{Gaunt factor}~\cite{gonoskov.rmp.2022}, is given by $g(\chi) = \moment(1,\chi)$.
Similarly, the function that controls variance growth due to stochasticity~\citep{ridgers.jpp.2017}, $g_2(\chi) = \moment(2,\chi)$.

The classical moments can be evaluated directly:
	\begin{equation}
	\mathcal{M}_\cl(n, \chi) =
		\frac{3^n \sqrt{3} \Gamma\!\left(\frac{n}{2}+\frac{1}{6}\right)
				\Gamma\!\left(\frac{n}{2}+\frac{11}{6}\right)}{2 \pi (n + 1)}
		\alpha m \chi^{2n + 1},
	\end{equation}
where the gamma function is defined by $\Gamma(z) = \int_0^\infty t^{z-1} e^{-z} \,\rmd z$.

\begin{widetext}
The quantum moments cannot be expressed in closed form, so it is more convenient to quote their normalized values.
The first step is to express $\moment$ as a single integral:
	\begin{equation}
	\moment(n,\chi) =
		\frac{2}{\Gamma\!\left(\frac{n}{2}+\frac{1}{6}\right)
					\Gamma\!\left(\frac{n}{2}+\frac{11}{6}\right)}
		\int_0^\infty\!
			\left[
			\frac{(n+1) y^n (8+12\chi y + 9\chi^2 y^2) K_{2/3}(y)}{(2 + 3\chi y)^{n+3}}
			- \frac{y^{n+1} K_{1/3}(y)}{(2 + 3\chi y)^{n+1}}
			\right]
		\rmd y,
	\label{eq:NormalizedMoment}
	\end{equation}
which can be evaluated numerically for any $n$ and $\chi$.
Limiting values of \cref{eq:NormalizedMoment} are, for $\chi \ll 1$,
	\begin{equation}
	\moment (n,\chi) =
		1 -
		\frac{3 (n+1)
				\Gamma\!\left(\frac{n}{2}+\frac{2}{3}\right)
				\Gamma\!\left(\frac{n}{2}+\frac{7}{3}\right)}{%
				\Gamma\!\left(\frac{n}{2}+\frac{1}{6}\right)
				\Gamma\!\left(\frac{n}{2}+\frac{11}{6}\right)
				} \chi
		+ \frac{(n+1)(3n+1)[28+n(3n+17)]}{8} \chi^2 + \cdots
	\end{equation}
and for $\chi \gg 1$,
	\begin{equation}
	\moment (n,\chi) =
		-\frac{(n+1) [28 + 9n(n+3)]
					\Gamma\!\left(-\frac{1}{3}\right)
					\Gamma\!\left(\frac{2}{3}\right)
					\Gamma\!\left(n+\frac{1}{3}\right)}{%
					27\,
					\Gamma\!\left(\frac{n}{2}+\frac{1}{6}\right)
					\Gamma\!\left(\frac{n}{2}+\frac{11}{6}\right)
					\Gamma\!\left(n+3\right)}
		(3\chi)^{-n-1/3}.
	\end{equation}
Examples of moments at specific orders are given in \cref{tbl:Results}.
\end{widetext}

\section{Mean energy loss}
\label{sec:Mean}

We begin by expanding \cref{eq:MuEvolution} to first order in $\chi_0$.
This requires $g(\chi)$ to first order in $\chi$, which is given in \cref{tbl:Results} in \cref{sec:Moments}:
    \begin{multline}
    \frac{\rmd \mf}{\rmd\phi} =
        -\frac{2}{3} R_c \abs{f(\phi)}^2 \mf^2
        \left[
            \left( 1 + \frac{\sigma^2}{\mu^2} \right)
    \right. \\ \left.
            \phantom{} - \frac{55 \sqrt{3}}{16} \chi_0 \abs{f(\phi)} \mf
            \left( 1 + \frac{3\sigma^2}{\mu^2} + \frac{\varsigma^3}{\mu^3} \right)
        \right]
    \label{eq:MeanEOM}
    \end{multline}
where $\mf = \mu / \mu_0$ is the mean energy normalized to its initial value.
The expansion in \cref{eq:MeanEOM} is effectively an expansion to first order in $\hbar$, even though we work in natural units, because
$R_c \propto \hbar^0$ by virtue of the factor of $\alpha$.
If the corrections due to the higher order moments $\sigma^2,\varsigma^3$ are subleading with respect to the quantum correction $\propto \chi_0 \mf$, we may neglect all terms containing those higher order moments and solve this perturbatively by introducing $\mf = \mf^{(0)} + \chi_0 \mf^{(1)} + O(\chi_0^2)$.
The result is
    \begin{multline}
    \mf(\phi) =
        \frac{1}{1 + \frac{2}{3} R_c I(\phi)}
    \\
            \phantom{} + \frac{55 \chi_0}{8 \sqrt{3} [1 + \frac{2}{3} R_c I(\phi)]^2}
            \int_{-\infty}^\phi\! \frac{R_c \abs{f(\psi)}^3}{1 + \frac{2}{3} R_c I(\psi)} \,\rmd\psi,
    \label{eq:Mean}
    \end{multline}
where $I(\phi) = \int_{-\infty}^\phi \abs{f(\psi)}^2 \rmd\psi$.
The first term is the classical result, where the total energy loss depends on the integrated flux~\citep{dipiazza.lmp.2008}.
The second term is positive, indicating that the total radiated energy is reduced~\citep{erber.rmp.1966}.

	\begin{figure}
	    \includegraphics[width=0.9\linewidth]{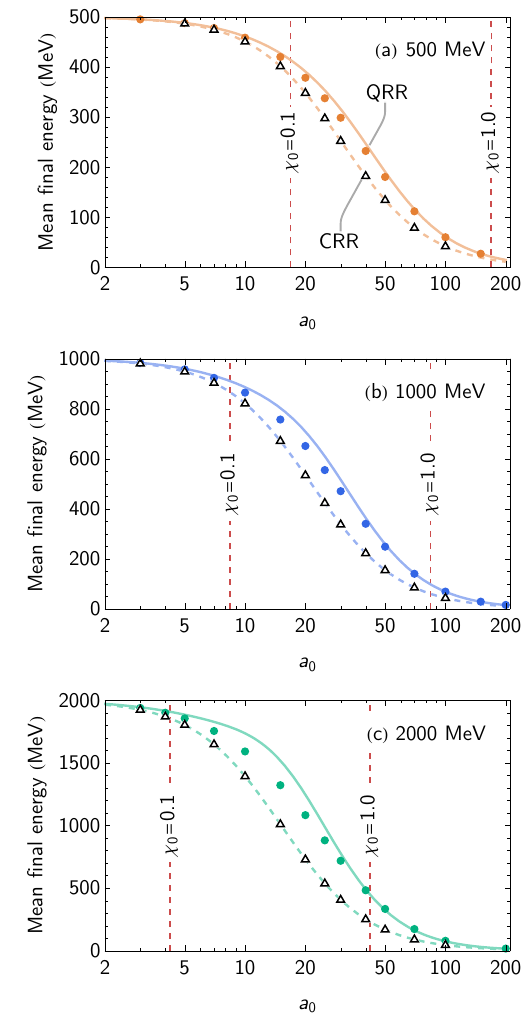}
	    \caption{%
            The mean final energy from simulations (points) and as predicted by \cref{eq:Mean}, for quantum (solid) and classical (dashed) radiation reaction.
            The electron beam is initialised with a mean energy of (a) 500 MeV, (b) 1000 MeV or (c) 2000 MeV.
            The laser pulse has a normalized amplitude of $a_0$, a wavelength of $0.8~\micron$ and a FWHM duration of 30~fs.
	    }
	\label{fig:MeanEnergy}
	\end{figure}

A comparison of \cref{eq:Mean} with the results of numerical simulations, performed with the Monte-Carlo particle-tracking code Ptarmigan v1.3.2~\cite{ptarmigan,blackburn.pop.2023}, is given in \cref{fig:MeanEnergy}.
In these simulations an electron beam with mean energy 500, 1000 or 2000 MeV (Gaussian energy distribution, with 10\% energy spread) collides with a plane-wave laser pulse with Gaussian temporal envelope, normalized amplitude $a_0$, a wavelength of $0.8~\micron$ and a FWHM duration of 30~fs.
We vary $a_0$ in the range $2 < a_0 < 200$ and use either a quantum (stochastic) model of radiation reaction, which builds on LCFA photon emission rates~\cite{erber.rmp.1966,ritus.jslr.1985}, or a classical model, which uses the Landau-Lifshitz equation~\cite{landau.lifshitz}.
One may see that the agreement is rather good across the full range of parameters, even though $\chi_0$ is not necessarily much smaller than unity.
This may be explained by the fact that our results are first-order in $\chi_0$, but ``all-order'' in the radiation-reaction parameter $R_c$:
as the electron beam loses energy, its instantaneous quantum parameter is reduced and so too the importance of quantum corrections (see Ref.~\citenum{popruzhenko.njp.2019} for a similar result).

It may be seen, however, that the theory generally underestimates the energy loss in the quantum case.
This is particularly visible for $E_0 = 2000$~MeV around $a_0 \simeq 15$.
We explain this by referring the reader to the neglect of higher-order moments in \cref{eq:MeanEOM}.
If the electron energy distribution is very broad ($\sigma \sim \mu$), the first term, which describes energy loss, is increased in magnitude.
This is not generally significant under classical radiation reaction, because the variance only ever decreases.
Under quantum radiation reaction, by contrast, stochastic effects make it possible for an initially monoenergetic electron beam to develop a broad energy spread.
It is reasonable to expect that the error made by \cref{eq:Mean} is largest for those collision parameters where the energy spread midway through the laser pulse is largest.
We turn therefore to the solution of \cref{eq:SigmaEvolution}, which describes how the variance of the energy distribution evolves.

\section{Broadening and narrowing of the energy spectrum}
\label{sec:Variance}

Expanding \cref{eq:SigmaEvolution} to first order in $\chi_0$, and neglecting moments of higher order than $\sigma^2$ for brevity, yields an equation of motion for the normalized variance $\hat{\sigma} = \sigma / \mu_0$:
    \begin{multline}
    \frac{\rmd \hat{\sigma}^2}{\rmd \phi} =
        -\frac{8}{3} R_c \abs{f(\phi)}^2 \mf \hat{\sigma}^2
        + \frac{55 \sqrt{3}}{4} R_c \chi_0 \abs{f(\phi)}^3 \mf^2 \hat{\sigma}^2
    \\
        + \frac{55}{4 \sqrt{3}} R_c \chi_0 \abs{f(\phi)}^3 \mf^2 \left( \hat{\sigma}^2 + \frac{\mf^2}{6} \right).
    \label{eq:VariancePhase}
    \end{multline}
In the classical limit $\chi_0 \to 0$, we have
    \begin{equation}
    \hat{\sigma}^2_\text{cl}(\phi) = \frac{\hat{\sigma}^2_0}{[1 + \frac{2}{3} R_c I(\phi)]^4},
    \label{eq:ClassicalVariance}
    \end{equation}
which can be expressed as $\sigma/\sigma_0 = (\mu/\mu_0)^2$ in agreement with \citet{neitz.prl.2013} and \citet{vranic.prl.2014}.
This could, in principle, be corrected for non-zero $\chi_0$ in much the same way as done for the mean energy loss, by expanding $\hat{\sigma}^2 = \hat{\sigma}^2_{(0)} + \chi_0 \hat{\sigma}^2_{(1)}$ where $\hat{\sigma}^2_{(0)}$ is the classical result in \cref{eq:ClassicalVariance}.
However, the quantum terms in \cref{eq:VariancePhase} are not necessarily small corrections to the classical terms.
Consider an initially monoenergetic beam, with $\hat{\sigma} = 0$: the leading order term in this scenario is the purely quantum term $\propto \mf^4$, which drives growth of the variance.
The first and second terms, which represent the reduction in the variance due to (quantum-corrected) radiation losses, do not dominate until $\hat{\sigma}$ has grown to a sufficiently large value.

Therefore we introduce a new parameter $V$, defined by $\hat{\sigma}^2 = \chi_0 V$, before perturbatively expanding in $\chi_0$, i.e. $V = V^{(0)} + \chi_0 V^{(1)}$.
The equation of motion for $V^{(0)}$ contains the first and last terms of \cref{eq:VariancePhase}, the competing growth and suppression, at the same order, as desired.
Solving this, and then writing $\hat{\sigma}^2 = \chi_0 V^{(0)}$, we obtain:
    \begin{multline}
    \hat{\sigma}^2_\text{q}(\phi) =
        \frac{1}{[1 + \frac{2}{3} R_c I(\phi)]^4}
    \\
        \times \left(
            \hat{\sigma}^2_0
            + \frac{55 R_c \chi_0}{24 \sqrt{3}}
            \int_{-\infty}^\phi \! \abs{f(\psi)}^3 \,\rmd\psi
        \right)
    \label{eq:Variance}
    \end{multline}
One can identify two regimes of behaviour in \cref{eq:Variance}:
in the first, the initial variance is sufficiently large that the stochastically driven growth is a small correction; and in the second, the radiation-loss-driven reduction in the variance is a small correction to the growth.
\Citet{niel.pre.2018} refer to these as the \emph{cooling} and \emph{heating} regimes respectively.

	\begin{figure}
	    \includegraphics[width=0.9\linewidth]{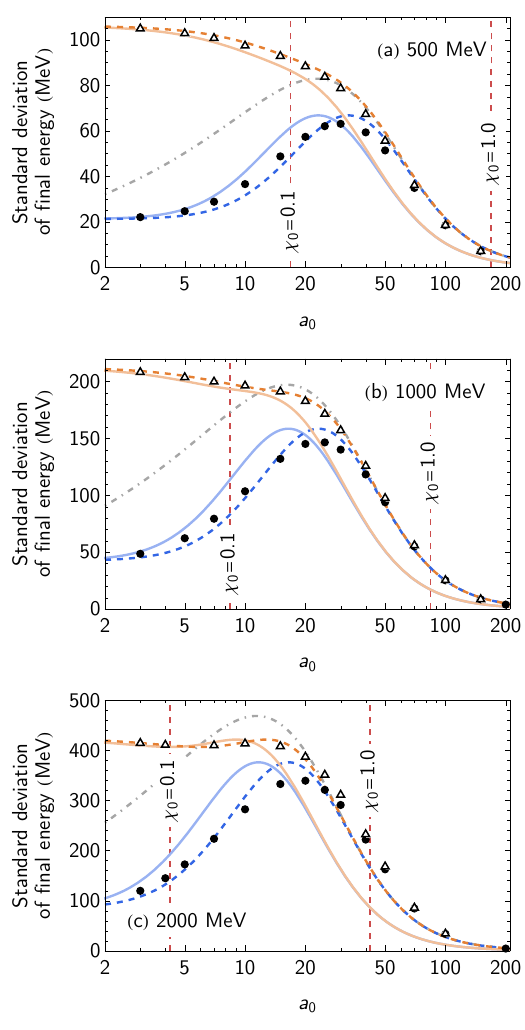}
	    \caption{%
            The standard deviation of the final energy from simulations (points), and as predicted by \cref{eq:Variance} (solid lines), for an electron beam with initial mean energy (a) 500 MeV, (b) 1000 MeV or (c) 2000 MeV undergoing quantum radiation reaction.
            Dashed lines give an ad hoc corrected \cref{eq:Variance} (see text for details).
            Grey, dot-dashed lines give Eq.~17 from \citet{vranic.njp.2016}.
            The electron beam is initialised with a Gaussian energy distribution, with FWHM equivalent to 10\% (solid disks) or 50\% (open triangles) of the mean energy.
            The laser pulse has a normalized amplitude of $a_0$, a wavelength of $0.8~\micron$ and a FWHM duration of 30~fs.
	    }
	\label{fig:StandardDeviation}
	\end{figure}

We now compare this prediction to the results of numerical simulations.
Here we consider the case of quantum radiation reaction and investigate the role of the initial variance $\sigma_0^2$.
The electron beam is initialized with a Gaussian energy distribution, with a mean of 500, 1000 or 2000 MeV and a spread (defined by the FWHM) of either 10\% or 50\% of the mean.
As before, the laser pulse is a plane wave with Gaussian temporal envelope, normalized amplitude $a_0$, a wavelength of $0.8~\micron$ and a FWHM duration of 30~fs.
Our results are given in \cref{fig:StandardDeviation}.
The qualitative agreement between the theory (solid lines) and simulation results (points) is reasonably good.
We see that if the initial energy spread is small, stochasticity drives broadening of the spectrum that is maximized at a particular $a_0$.
However, if the $a_0$ is increased beyond this point, radiative cooling dominates and the energy spread is reduced.
If, on the other hand, the initial energy spread is large, no stochastic broadening is visible.

The quantitative agreement is not as good as found for the mean energy, because the contribution of higher-order moments is generally more important for the evolution of $\sigma^2$.
(Stochastic broadening leads to increases in both the variance and the skewness, for example~\cite{niel.pre.2018}.)
However, this may be improved significantly by scaling $a_0 \to a_0 / \sqrt{2}$ in \cref{eq:Variance}.
With this correction, shown by the dashed lines in \cref{fig:StandardDeviation}, the agreement is good across the full range of $a_0$.
The effectiveness of this ad hoc approach may be explained by the fact that it reduces the cooling, which \cref{eq:VariancePhase} overestimates because it contains no higher-order moments.

\Cref{fig:StandardDeviation} also shows the standard deviation predicted by Eq.~17 in \citet{vranic.njp.2016}, which is derived under the assumptions that the initial energy spread is small and that laser pulse is long enough that the variance has grown to its maximal value before beginning to shrink.
This scaling law is in excellent agreement with our simulation results if $a_0$ is large, where these assumptions are valid: both \cref{eq:ApproxVariance} and Eq.~17 in \cite{vranic.njp.2016} predict that $\sigma^2_\mathrm{f} \propto a_0^{-5}$ if $a_0 \gg 1$.
It is less accurate for intermediate $a_0$, where stochastic broadening and radiative cooling are comparable in magnitude, or if the initial energy spread is large.

\section{Discussion}

Here we present \cref{eq:Mean,eq:Variance} in a more practical form.
We consider the case of a linearly polarized laser pulse with a Gaussian temporal envelope, for which $\vec{f}(\phi) = \vec{e}_x \sin\phi \, \exp(-2 \ln 2 \, \phi^2 / \tau^2)$.
Assuming further that the phase duration $\tau \gg 2\pi$, we may average over the fast oscillations and obtain $I(\phi) = (\tau/8) \sqrt{\pi / \ln 2} \, [1 + \text{erf}(2 \sqrt{\ln 2} \, \phi / \tau)]$.
The integral in \cref{eq:Mean} cannot be performed analytically: however, it may be shown to be a function of the single parameter $R_c \tau$, so we evaluate it numerically for various $R_c \tau$ and find a suitable fitting function.
Under quantum radiation reaction, the final (normalized) mean and variance are:
    \begin{align}
    \begin{split}
    \mf_\mathrm{f} &=
        \frac{1}{1 + 0.355 \, R_c \tau} 
        \left[
            1 +
            \frac{3.969 \, \chi_0 \, \mathcal{F}(R_c \tau)}{1 + 0.355 \, R_c \tau}
        \right],
    \\
    \mathcal{F}(R_c \tau) &= \frac{0.369 \, R_c \tau}{1 + 0.171 (R_c \tau)^{3/5} + 0.0819 \, R_c \tau},
    \end{split}
    \label{eq:ApproxMean}
    \end{align}
and
    \begin{equation}
    \hat{\sigma}^2_\mathrm{f} =
        \frac{\hat{\sigma}^2_0 + 0.173 \, \chi_0 R_c \tau}{[1 + 0.178 \, R_c \tau]^4},
    \label{eq:ApproxVariance}
    \end{equation}
where we have included the the ad hoc correction discussed in \cref{sec:Variance}.
Under classical radiation reaction, we have instead $\mf_\mathrm{f} = (1 + 0.355 R_c \tau)^{-1}$ and $\hat{\sigma}^2_\mathrm{f} = \hat{\sigma}_0^2 / (1 + 0.355 R_c \tau)^4$.
The collision parameters are given by:
    \begin{align}
    \begin{split}
    \chi_0 &= 0.812 \left( \frac{E_0}{\text{GeV}} \right) \left( \frac{I_0}{10^{22}~\text{W}\text{cm}^{-2}} \right)^{1/2},
    \\
    \tau &= 1.85 \left( \frac{T}{\text{fs}} \right)  \left( \frac{\lambda}{\mu\mathrm{m}} \right)^{-1},
    \\
    R_c \tau &= 0.954 \left( \frac{E_0}{\text{GeV}} \right) \left( \frac{I_0}{10^{22}~\text{W}\text{cm}^{-2}} \right) \left( \frac{T}{\text{fs}} \right),
    \end{split}
    \end{align}
where $E_0$ is the mean initial energy of the electrons, $I_0$ is the laser intensity, $T$ is the full-width-at-half-maximum duration of the pulse intensity profile, and $\lambda$ is the laser wavelength.

\begin{figure}
    \includegraphics[width=0.8\linewidth]{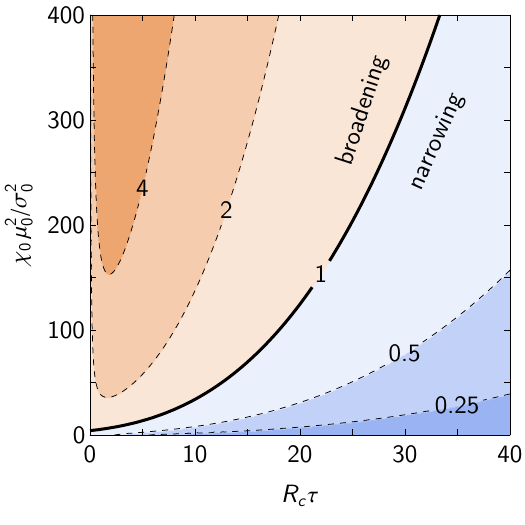}
    \caption{%
        The ratio of the final and initial standard deviations, $\sigma_\mathrm{f} / \sigma_0$, predicted by the corrected \cref{eq:ApproxVariance}, with broadening ($\sigma_\mathrm{f} > \sigma_0$) in orange and narrowing ($\sigma_\mathrm{f} < \sigma_0$) in blue.
        Contour lines indicate where $\sigma_\mathrm{f} / \sigma_0$ is equal to the labelled value.
    }
\label{fig:Broadening}
\end{figure}

Let us consider what these results imply about the collision parameters under which stochastic broadening may be expected.
We see from \cref{eq:ApproxVariance} that the ratio of the final and initial standard deviations, $\sigma_\mathrm{f} / \sigma_0$, is a function of two parameters: a scaled quantum parameter $\chi_0 \mu_0^2 / \sigma_0^2$ and a duration-weighted radiation-reaction parameter $R_c \tau$.
The region in which stochastic broadening overcomes both the initial energy spread and the effect of radiative cooling is indicated in orange in \cref{fig:Broadening}.
It is accessed by increasing the quantum parameter and reducing the initial variance.
By contrast, an increase in $R_c \tau$ is generally associated with an increase in radiation losses, which eventually reduce the energy spread.
Differentiating \cref{eq:ApproxVariance} with respect to $R_c \tau$ reveals that there is a maximum at positive $R_c \tau$ if $\sigma_0^2 \lesssim 0.25 \chi_0 \mu_0^2$, namely $\max (\sigma_\mathrm{f}^2) \simeq 0.10 \chi_0 \mu_0^2 / [1 - \sigma_0^2 / (\chi_0 \mu_0^2)]^3$.
This is in reasonable agreement with the maximum energy spread (the `turning point'~\cite{vranic.njp.2016} or `threshold variance'~\cite{niel.pre.2018}) calculated by \citet{vranic.njp.2016} and \citet{niel.pre.2018}.

The competition between these factors means that stochastic broadening is maximized at a particular $a_0$~\cite{arran.ppcf.2019}, which we derive from \cref{eq:ApproxVariance} under the assumption that the initial variance is small and all other quantities are held constant:
    \begin{equation}
    a^\text{opt}_0 \simeq 160 \left( \frac{E_0}{\text{GeV}} \right)^{-1/2} \left( \frac{T}{\text{fs}} \right)^{-1/2}  \left( \frac{\lambda}{\mu\mathrm{m}} \right).
    \label{eq:Optimala0}
    \end{equation}
The value of the standard deviation at the given optimum is:
    \begin{equation}
    \sigma_\mathrm{f}  = 370 \left( \frac{E_0}{\text{GeV}} \right)^{5/4} \left( \frac{T}{\text{fs}} \right)^{-1/4} \text{MeV}.
    \end{equation}
These predict that $a^\text{opt}_0 = \{33, 23, 17\}$ and $\sigma_\mathrm{f} = \{ 67, 160, 380 \}$~MeV for initial energies of $\{0.5, 1, 2\}$~GeV, which is consistent with the results shown in \cref{fig:StandardDeviation}.
By expressing $\sigma_\mathrm{f} / \sigma_0$ as a function of $\ln a_0$ and expanding around $\ln a_0^\text{opt}$ to second order, we can also estimate the width of this maximum to be $(1/3) a_0^\text{opt} \lesssim a_0 \lesssim 3 a_0^\text{opt}$; this too is consistent with \cref{fig:StandardDeviation}.

It is important to bear in mind that the results in this work have been derived for plane-wave laser pulses.
Since the electron beam and laser pulse in a real experiment are likely to have comparable transverse dimensions ($\sim \micron$), finite-size effects are significant.
Effectively this means that the different components of the electron beam `see' different peak intensities.
The relevant signals are then integrated over a distribution of effective $a_0$, $dN_e/da$, where $0 < a < a_0$:
\edit{it complicates the identification of quantum radiation reaction effects if the laser pulse and electron beam have comparable transverse sizes (see analysis in \citet{poder.prx.2018}).
Indeed}, broadening of the electron energy distribution would be expected even under classical radiation reaction.
The question of whether stochastic broadening is still observable, despite finite-size effects, can be approached directly using 3D simulations.
On the other hand, \citet{amaro.njp.2021,amaro.arxiv.2023} have shown that plane-wave scaling laws, such as those we have here, can be adapted to the fully 3D situation by considering the structure of $dN_e/da$.

	\begin{figure}
	    \includegraphics[width=0.9\linewidth]{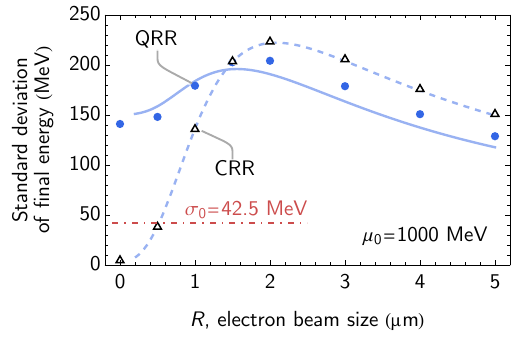}
	    \caption{%
            The final standard deviation in energy of a 1-GeV electron beam (10\% energy spread, transverse size $R$) that collides with a 30-fs laser pulse that is focused to $w_0 = 2.5~\micron$ and $a_0 = 30$: from (points) simulations and \cref{eq:Variance3d} for (solid line) quantum and (dashed line) classical radiation reaction.
	    }
	\label{fig:3d}
	\end{figure}

The mean and variance that characterize a beam of electrons are
    \begin{align}
    \hat{\mu}_\text{f,b} &= \frac{1}{N_e} \int_0^{a_0} \! \frac{d N_e}{d a} \hat{\mu}_\mathrm{f}(a) \, \rmd a,
    \\
    \hat{\sigma}^2_\text{f,b} &= \frac{1}{N_e} \int_0^{a_0} \! \frac{d N_e}{d a} \left[ \hat{\mu}^2_\mathrm{f}(a) + \hat{\sigma}^2_\mathrm{f}(a) \right] \, \rmd a - \hat{\mu}^2_\text{f,b},
    \label{eq:Variance3d}
    \end{align}
where we emphasise that $\hat{\mu}_\mathrm{f}$ and $\hat{\sigma}_\mathrm{f}$, i.e. \cref{eq:ApproxMean,eq:ApproxVariance}, are functions of the effective amplitude $a$.
Let us consider an electron beam with spherically symmetric, Gaussian charge density (rms size $R$) that collides with a focused laser pulse with waist $w_0$.
Assuming that $R$ is much smaller than the laser Rayleigh range, and that there is no transverse displacement between the beams, we have~\cite{vranic.njp.2016}:
    \begin{equation}
    \frac{d N_e}{d a} = \frac{N_e w_0^2}{a R^2} \exp\left(\frac{w_0^2}{R^2} \ln \frac{a}{a_0}\right).
    \end{equation}
As an example, we compare these 3D-weighted scaling laws with simulations in \cref{fig:3d}, for $a_0 = 30$, $w_0 = 2.5~\micron$, $\lambda = 0.8~\micron$, $T = 30$~fs, $\mu_0 = 1000$~MeV and $\sigma_0$ equivalent to 10\% energy spread.
This set of collision parameters is close to the optimum identified in \cref{eq:Optimala0} (see also \cref{fig:Broadening}).
We find not only good agreement between the theory and simulations, but that broadening occurs in both the classical and quantum cases.
The two can be distinguished, and specifically stochastic effects observed, only if the transverse size of the electron beam is smaller than the laser waist.

\section{Summary}

We have presented analytical predictions for the mean and variance of the energy distribution of electron beams that collide with high-intensity laser pulses.
This work extends results obtained earlier for classical radiation reaction~\cite{dipiazza.lmp.2008,neitz.prl.2013,vranic.prl.2014,vranic.njp.2016} to the quantum regime.
Despite the fact our results are derived assuming that the quantum parameter $\chi_0$ is small, we find that they give accurate predictions for parameters relevant for upcoming experiments, namely $a_0 < 200$ and initial electron energies in the GeV range.
In particular, we are able to show how the initial energy spread of the electron beam affects the possibility to observe stochastic broadening.
\edit{As it focuses on statistical measures of the electron spectrum, this work will be relevant for upcoming experiments, which will achieve many more collisions at high intensity than were obtained in the first experimental campaigns~\cite{cole.prx.2018,poder.prx.2018}.}

From our analysis it may be concluded that the best approach to experimental observation of stochastic broadening is to optimize the energy spread and stability of the electron beam, rather than pushing towards higher intensity or electron-beam energy.
Increasing the laser intensity in particular is likely to be counterproductive, as it enhances radiative cooling ($R_c \propto a_0^2$) more than it increases the quantum parameter ($\chi \propto a_0$).
The scaling laws presented here indicate that a conclusive observation of quantum radiation reaction is well within the capability of current high-intensity laser facilities.

\begin{acknowledgments}
The author thanks Stepan Bulanov, Caterina Riconda, Christopher Ridgers, Greger Torgrimsson and Marija Vranic for useful discussions.
\end{acknowledgments}

\section*{Data availability}

Simulation results were obtained with Ptarmigan v1.3.2~\cite{blackburn.pop.2023}, available at \url{https://doi.org/10.5281/zenodo.7974876}, and may be reproduced using the Supplemental Material.

\bibliography{references}

\end{document}